\documentclass[twocolumn,preprintnumbers,amsmath,amssymb,superscriptaddress]{revtex4}

\usepackage{graphicx}
\usepackage{dcolumn}
\usepackage{bm}

\begin{document}

\preprint{preprint}

\title{Microwave band on-chip coil technique\\ for single electron spin resonance in a quantum dot}

\author{Toshiaki Obata}
\author{Michel Pioro-Ladri{\`e}re}
\author{Toshihiro Kubo}
\author{Katsuharu Yoshida}
\affiliation{%
Quantum Spin Information Project, ICORP,JST,
Atsugi-shi, Kanagawa 243-0198, Japan
}%

\author{Yasuhiro Tokura}
\affiliation{%
Quantum Spin Information Project, ICORP,JST,
Atsugi-shi, Kanagawa 243-0198, Japan
}%
\affiliation{
NTT Basic Research Laboratories, NTT Corporation, 
Atsugi-shi, Kanagawa 243-0198, Japan
}

\author{Seigo Tarucha}

\affiliation{%
Quantum Spin Information Project, ICORP,JST,
Atsugi-shi, Kanagawa 243-0198, Japan
}%
\affiliation{
Department of Applied Physics, University of Tokyo,
Hongo, Bunkyo-ku, Tokyo 113-0033, Japan
}%

\date{\today}

\begin{abstract}
Microwave band on-chip microcoils are developed for the application to single electron spin resonance measurement with a single quantum dot.  
Basic properties such as characteristic impedance and electromagnetic field distribution are examined for various coil designs by means of experiment and simulation.  
The combined setup operates relevantly in the experiment at dilution temperature.  
The frequency responses of the return loss and Coulomb blockade current are examined.  
Capacitive coupling between a coil and a quantum dot causes photon assisted tunneling, whose signal can greatly overlap the electron spin resonance signal.  
To suppress the photon assisted tunneling effect, a technique for compensating for the microwave electric field is developed.  
Good performance of this technique is confirmed from measurement of Coulomb blockade oscillations.  
\end{abstract}

\maketitle

\section{\label{sec:Introduction}Introduction}

Electronic spin confined to quantum dots (QDs) is a robust quantum number \cite{Petta05} and is therefore regarded as a good candidate for forming a quantum bit in quantum information processing \cite{Hatano04,Hatano05}.  
Electron spin resonance (ESR) is a powerful tool for achieving coherent spin rotation in the qubit operation and is realized by applying a microwave (MW) magnetic field perpendicular to a static Zeeman field.  
ESR is usually achieved using a MW cavity.  
However, this technique is unsuitable for QDs since the MW cavity usually raises the temperature of the spin qubit to an unacceptable level because it transmits direct radiation from room temperature.  
One solution is to use a low temperature cavity.  
Then we deal with this heating problem by using an on-chip MW coil \cite{Kodera04,Kodera05,Wilfred06review} to which the MW is transmitted through coaxial cables to avoid direct room temperature radiation.  

Single spin ESR was recently demonstrated using a spin blockade readout technique for a double QD system with an on-chip coil structure \cite{Koppens06}.  
Unlike a single dot, the spin blockade charge readout scheme is efficient for realizing a low Zeeman field, in that it allows the on-chip coil to be operated at a low frequency (100 MHz).  
Here we aim to extend the frequency band to the gigahertz level, so a higher Zeeman field can be used that has much higher energy of $\sim$ 1 K (Refs. \cite{Koppens06} and \cite{LateralDotZeeman}) than the electron temperature.  

We set a static magnetic field to produce Zeeman splitting for an electron in a QD and apply MW.  
If the frequency matches the resonance condition, the MW magnetic field flips the spin to the excited state, which is positioned in the reservoir transport window.  
The electron can then exit the QD generating a finite current.  
However, capacitive coupling between a coil and a QD concurrently causes photon assisted tunneling (PAT)\cite{Kouwenhoven94PAT,WilfredPATrev,Wilfred03}, which also excites the electron out of the QD.  
This PAT process smears out the ESR signal.  
Therefore, it is crucial to minimize the MW electric field by optimizing the on-chip MW coil to suppress the PAT effect.  
We perform this on-chip coil optimizations after numerical simulation and experimental test.  

This paper is organized as follows.  
Section \ref{sec:theory} describes on-chip coil designs.  
The coil is connected to an on-chip high frequency transmission line to minimize insertion loss.  
Simulation results for several coil-waveguide structures are presented and compared.  
Transport measurements obtained through QD devices under MW excitation provided by the integrated coil are presented in Sec. \ref{sec:experiments}.  
We check the return loss, which affects the QD transport regarded as the PAT process.  
Koppens {\it et al.} \cite{Koppens06} previously developed a technique for canceling the PAT effect by using antiphase radio wave irradiation from a nearby antenna for lower frequencies.  
They attenuated and delayed the cancelation signal to minimize the MW electric field.  
The observed PAT signal was suppressed periodically with respect to frequency.  
We modified their technique for the higher frequency band using a variable phase shifter instead of a delay cable for tunable frequency.  
In Sec. \ref{sec:discussion}, we analyze the experimental data and calculate the amplitude of the MW electric field.  
We compare them with the help of numerical simulation to finally estimate the MW magnetic field.  

\section{\label{sec:theory}Design of on-chip coil}

\subsection{Transmission line}

\begin{figure}
\begin{minipage}{0.12\textwidth}
Microstrip \\
\includegraphics[width=\textwidth]{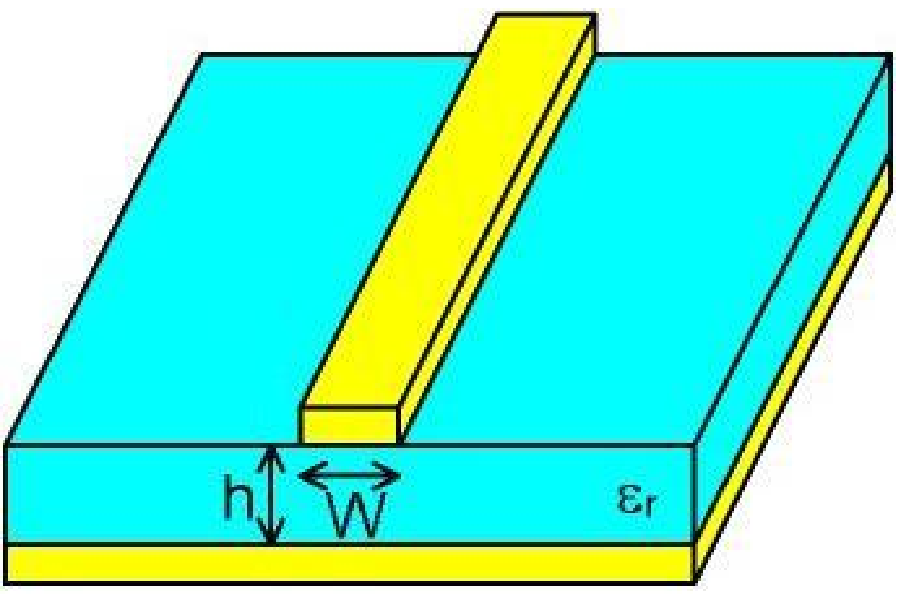}
\end{minipage}
\begin{minipage}{0.35\textwidth}
{\small
$
Z_0  = \frac{{30}}{{\sqrt {\varepsilon ' } }}\ln \left[ {1 + \frac{{4h}}{W}\left\{ {\frac{{8h}}{W} + \sqrt {\left( {\frac{{8h}}{W}} \right)^2  + \pi ^2 } } \right\}} \right]
$\\
$\varepsilon'=\frac{\varepsilon_r+1}{2} + \frac{\varepsilon_r-1}{2}/\sqrt{1+\frac{10h}{W}}$
}
\end{minipage}
\\
\vspace{0.1in}
\begin{minipage}{0.12\textwidth}
Coplanar strip (CPS)\\
\includegraphics[width=\textwidth]{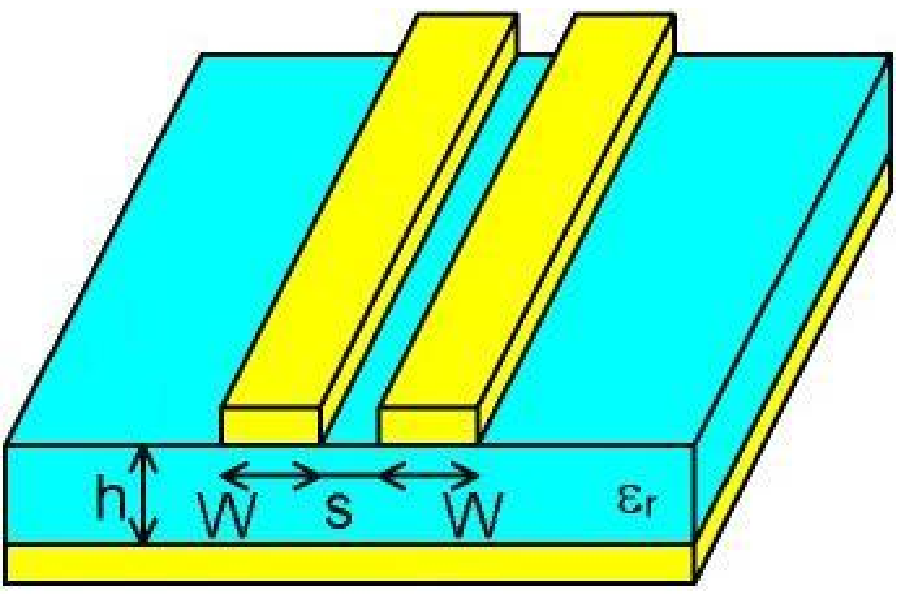}
\end{minipage}
\begin{minipage}{0.35\textwidth}
{\small
$
Z_0  = \frac{1}{{\sqrt {\varepsilon_{\it cop}} }}\sqrt {\frac{{\mu _0 }}{{\varepsilon _0 }}} \frac{{K\left( {s/(W + 2s)} \right)}}{{K\left( {\sqrt {1 - \left( {s/(W + 2s)} \right)^2 } } \right)}}
$\\
}
\end{minipage}
\\
\vspace{0.1in}
\begin{minipage}{0.12\textwidth}
Coplanar wave guide (CPW)\\
\includegraphics[width=\textwidth]{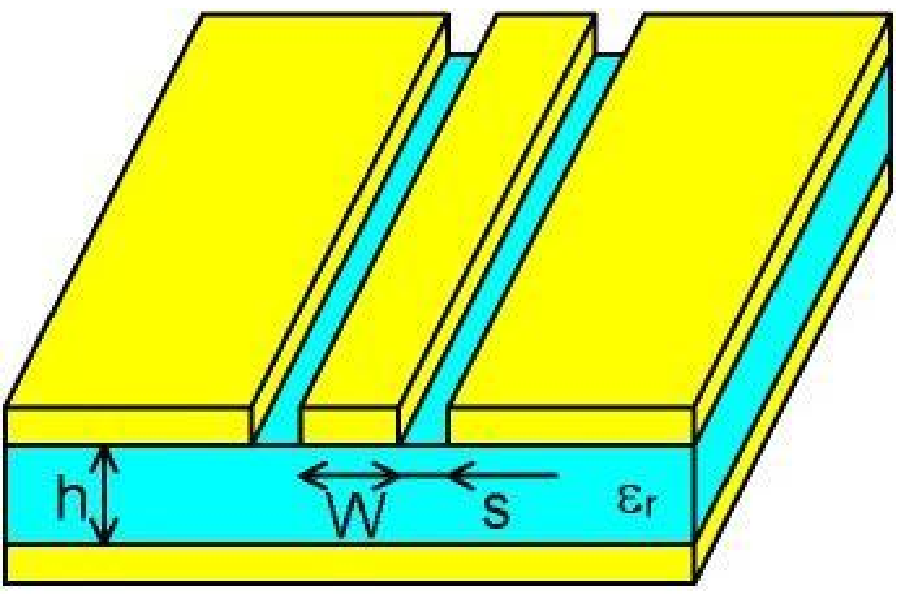}
\end{minipage}
\begin{minipage}{0.35\textwidth}
{\small
$
Z_0  = \frac{1}{{\sqrt {\varepsilon_{\it cop}} }}\frac{1}{4}\sqrt {\frac{{\mu _0 }}{{\varepsilon _0 }}} \frac{{K\left( {\sqrt {1 - \left\{ {W/(W + 2s)} \right\}^2 } } \right)}}{{K\left( {W/(W + 2s)} \right)}}
$\\
$\varepsilon_{\it cop}=1+\frac{\varepsilon _r - 1}{2}\cdot \frac{K(k')}{K(k)} \cdot \frac{K(k_1)}{K(k_1')}$\\
$k = \frac{W}{W + 2s}, k_1=\frac{\sinh(\pi W / 4h)}{\sinh(\pi(W + 2s) / 4h)}$\\
$k'=\sqrt{1-k^2}, k_1' = \sqrt{1 - k_1^2}$
}
\end{minipage}
\caption{\label{fig:onchiplines} Characteristic impedance of typical transmission lines.  $K$ is the first order complete elliptic integral.  $\mu_0$, $\varepsilon$, W, s, and h are the magnetic susceptibility, the dielectric constant, the metal width, the gap, and the substrate thickness, respectively.  
}
\end{figure}

\begin{table}
\caption{\label{table:matching} Propagation lines and the typical dimensions for the condition $Z_0$ = 50 $\rm\Omega$.  The physical parameters used for the calculation are as follows: metal thickness = 1 $\rm\mu m$, substrate thickness = 0.5 $\rm mm$, and dielectric constant of the substrate = 12.4 (Refs. \cite{GaAsepsilonMW} and \cite{GaAsepsilon}).  }
\begin{tabular}{ccc}
\hline
\hline
Lines & Metalwidth ($\rm \mu m$)&Gap ($\rm \mu m$)\\
\hline
Microstrip&360&...\\
\hline
Coplanar wave guide(CPW)&14&10\\
\hline
Coplanar strip(CPS)&100&10\\
\hline
\hline
\end{tabular}
\end{table}

In this section, we describe the design of our on-chip coil.  
The emerging problem in the MW band is heating.  
The metal becomes more resistive in proportion to the square root of the frequency because of the skin effect.  
We can deposit a several times thicker (approximately micrometer) metallic coil using a technique of photolithography rather than electron beam lithography.  
To obtain an ideal structure in terms of characteristic impedance, it is better to properly make for the metal thicker than 1 $\rm \mu m$.  
The typical skin depth decreases in inverse proportion to the square root of the frequency and is typically half a micrometer at 10 GHz.  
The line becomes more resistive than the simulation result for the metal thinner than 1 $\rm \mu m$.  
Thermal conductivity is larger for the thicker metal and thus allows the application of a MW signal with the larger current.  
The coil metal is approximately micrometer wide.  

The approximately micrometer wide coil is placed solely just around the QD.  
On-chip transmission lines are used to deliver a MW to the coil.  
The characteristic impedance $Z_0$ of these lines is adjusted to that of factory made coaxial cables (i.e., $Z_0$ = 50 $\rm\Omega$).  
Characteristic impedance is a function of the line geometry.  
According to textbooks such as Ref. \cite{okadabook04}, the formulas of $Z_0$ as a function of the dimensions for typical planar transmission lines are shown in Fig. \ref{fig:onchiplines}.  
The parameters are the metal width and gap.  
Their typical values are given in Table \ref{table:matching} for $Z_0$ = 50 $\rm\Omega$.  
The dielectric constant of the host material, GaAs, $\varepsilon _r =12.4$, is used for the calculation.  
The microstrip shown in Fig. \ref{fig:onchiplines} is the simplest for the MW transmission line.  
However, we did not employ it because it requires a width of as much as 360 $\rm\mu m$.  
Koppens {\it et al.} \cite{Koppens06} previously used a coplanar strip (CPS) transmission line for the ESR experiment.  
We used a coplanar waveguide (CPW) structure instead because the necessary surface area is smaller than CPS for an identical gap.  

Another factor related to the impedance is the reflection property.  
When the coil impedance $Z_L$ is connected to a transmission line with characteristic impedance $Z_0$, the MW is reflected at the input port of the coil.  
The reflection coefficient is $\rho = (Z_0 - Z_L) / (Z_0 + Z_L)$.  
In the present MW range, many parameters such as line curvature, resistivity, and metal thickness change the impedance.  
Bonding wire produces an additional impedance, which cannot be calculated easily.  
So, in this paper, we only examine if PAT occurs effectively (ineffectively) at the local minimum (maximum) reflection frequency (see Sec. \ref{sec:experiments}).  

\subsection{\label{sec:simulation}Numerical calculations}

\begin{table}
\caption{\label{table:simresults} Numerical estimations of electric and magnetic fields at center.  The excitation is a 1 V amplitude MW signal at 20 GHz. }
\begin{tabular}{ccc}
&Magnetic field (mT)&Electric field (mV/$\rm\mu m$)\\
\hline
\hline
Single&1.9&21\\
Spiral&0.5&4\\
Resonator&1.9&11\\
\hline
\hline
\end{tabular}
\end{table}

%
\begin{figure}
(a)\includegraphics[width=0.3\textwidth]{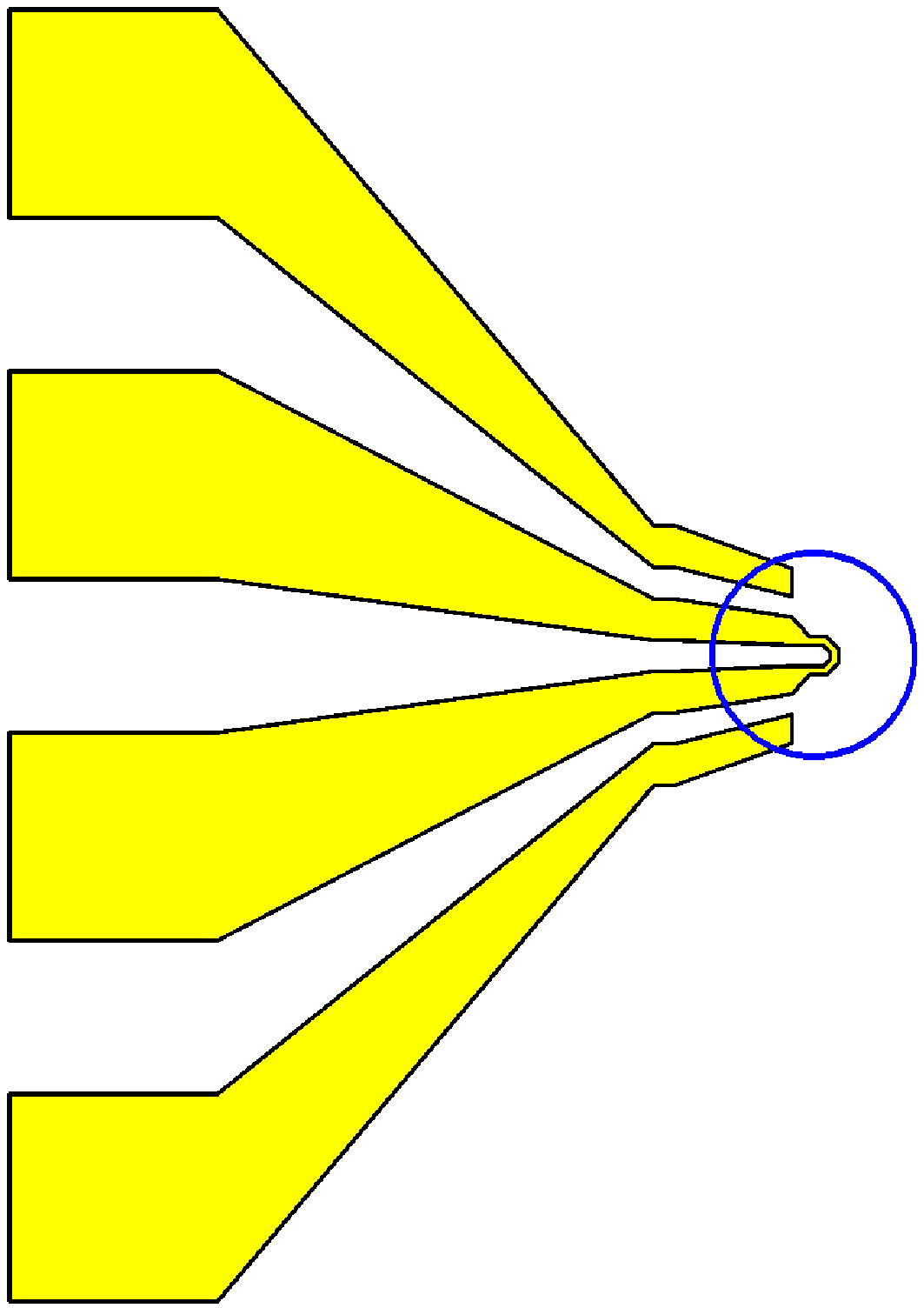}\\
(b)\includegraphics[width=0.3\textwidth]{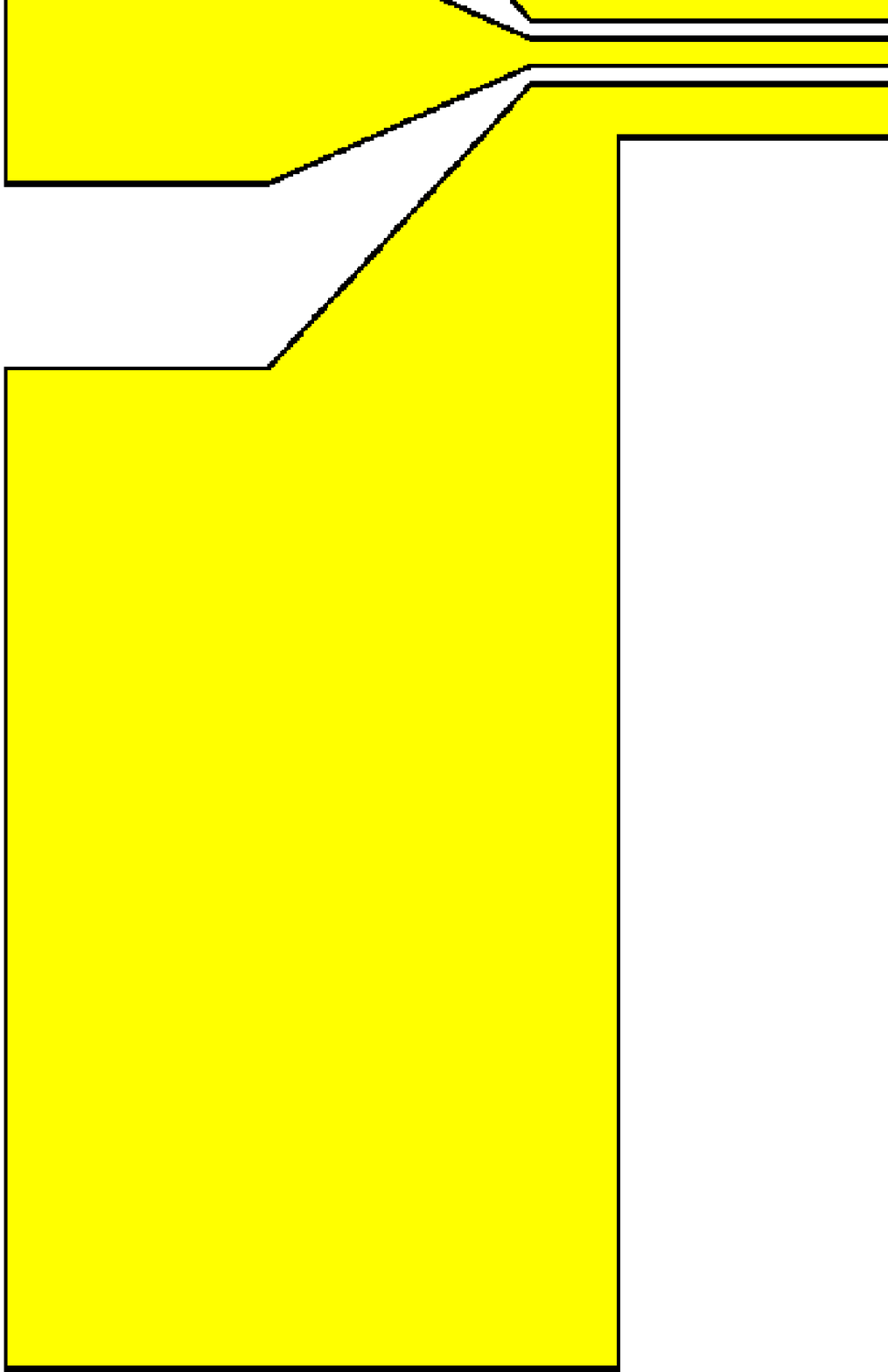}\\
(c)\includegraphics[width=0.3\textwidth]{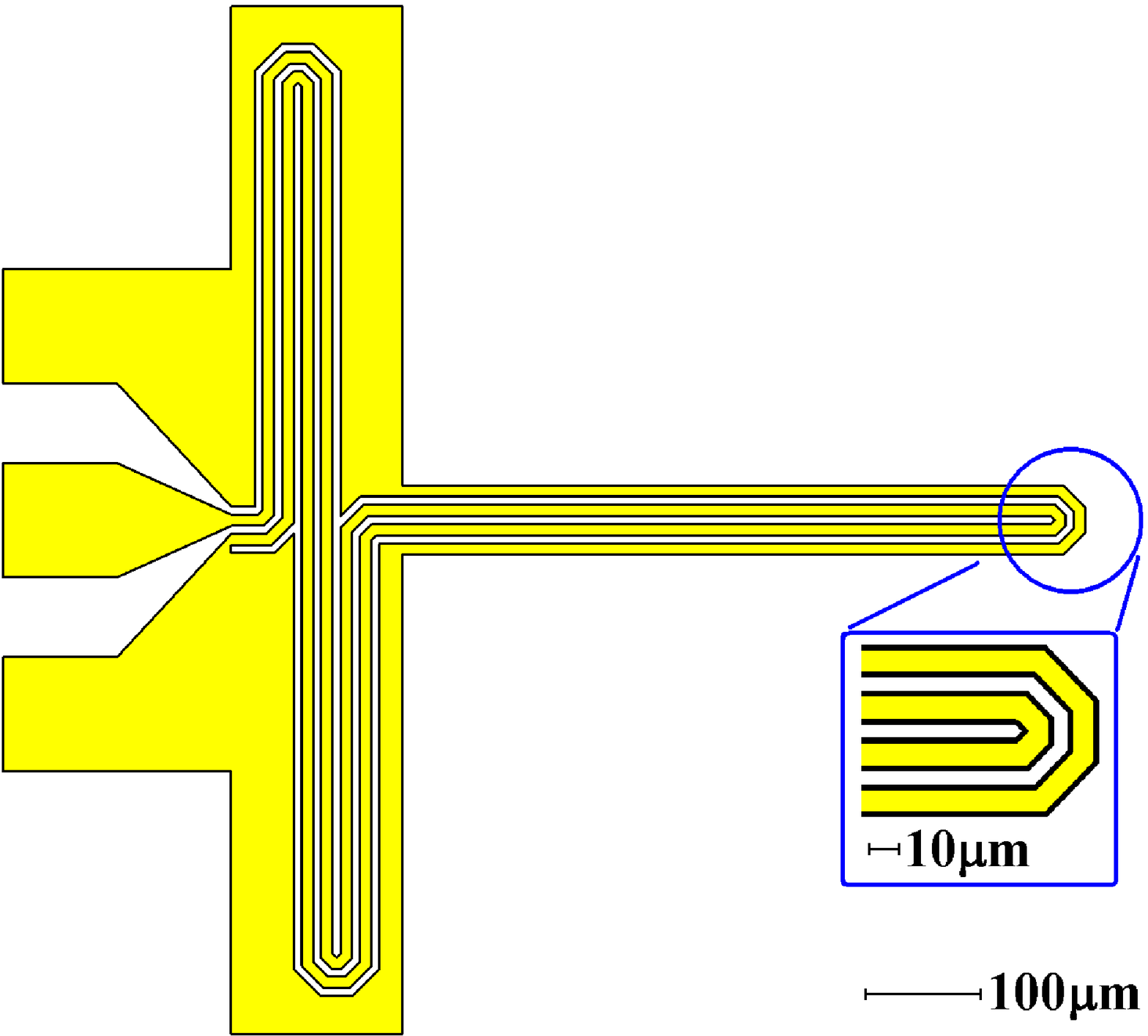}\\
\caption{\label{fig:single} CAD drawings of various on-chip coils:  (a) Single turn coil, (b) spiral coil, and (c) on-chip resonator.  Each coil is placed on high-dielectric substrate (GaAs).  The spiral coil has a three-dimensional structure, in which the current input at the left port flows into the coil center and to the right port through a bridge line (colored by red) isolated by 100 nm SiO$_2$ layer.  A quantum dot is placed at the center of the circle in each design.  }
\end{figure}

\begin{figure}
(a)\includegraphics[width=0.35\textwidth]{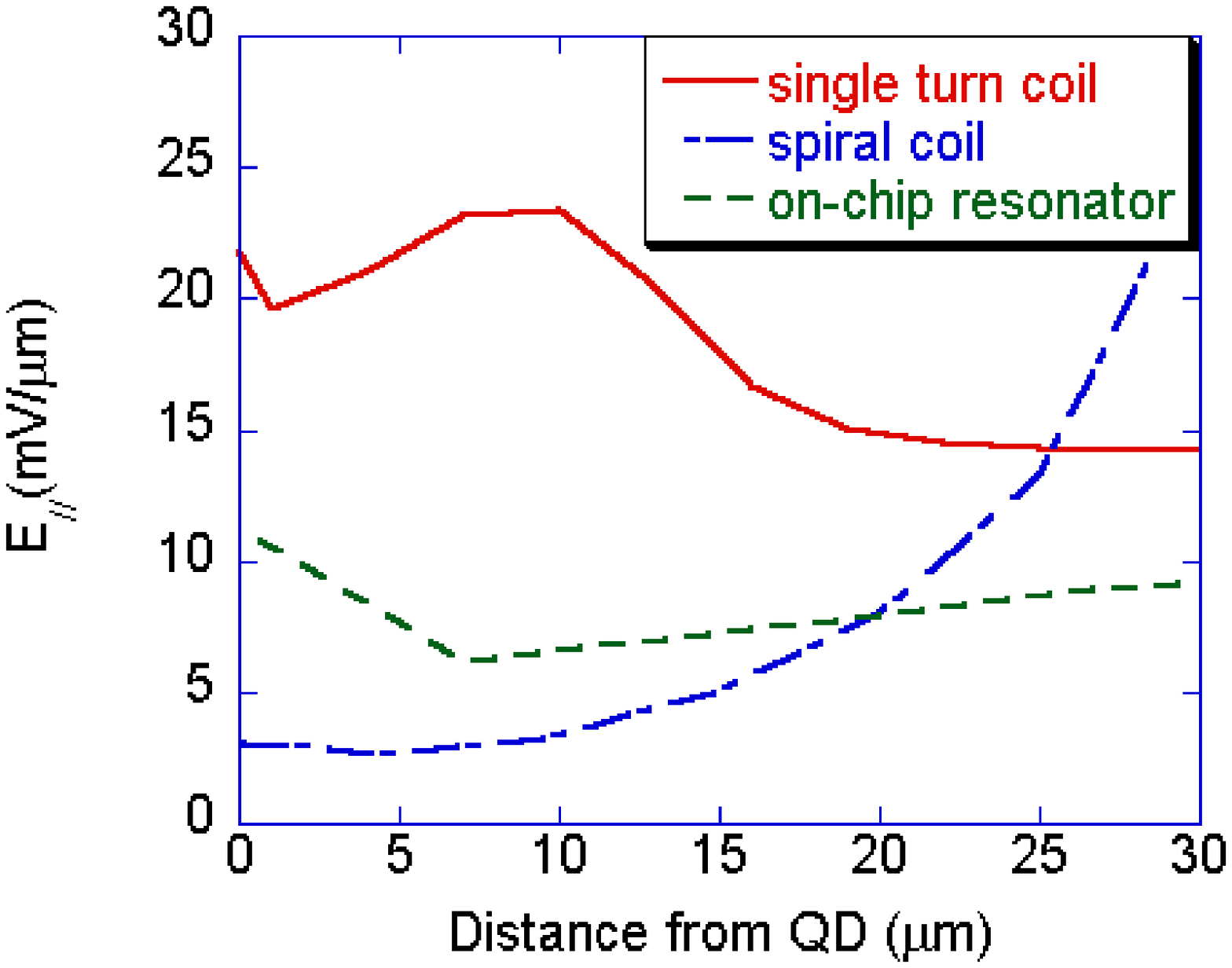}
(b)\includegraphics[width=0.35\textwidth]{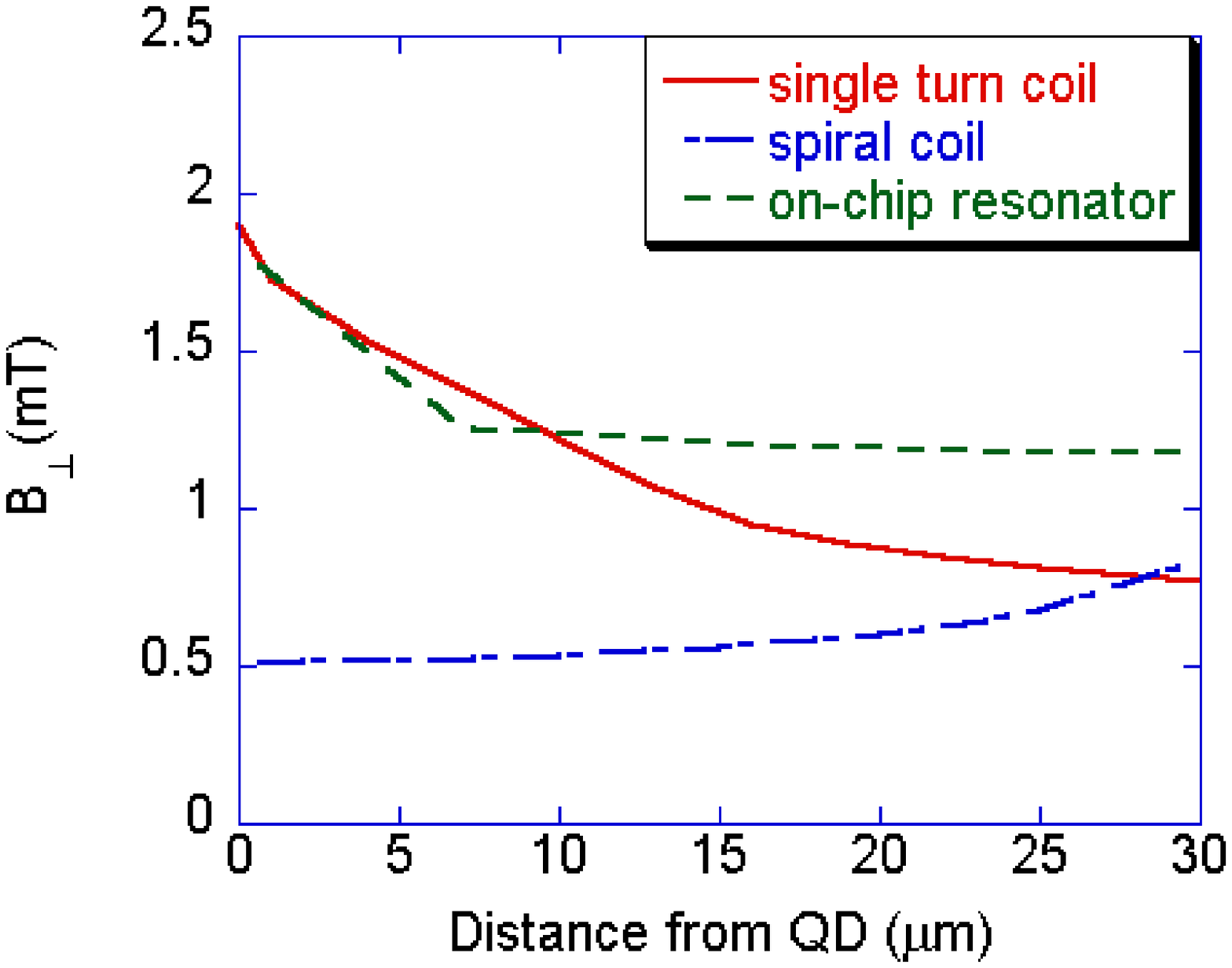}
\caption{\label{fig:MWfield}Calculation of MW electric (magnetic) field parallel (perpendicular) to the substrate (a) and (b), respectively.  The distance is measured horizontally from the center of the coil circle to the left in Fig. \ref{fig:single}.  The excitation is a 1 V amplitude MW signal at 20 GHz.  }
\end{figure}

We start with a simple design, which involves winding a CPW line to form a single turn coil [Fig. \ref{fig:single}(a)].  
When a CPW is bent at 180$^\circ$, a ground plane always remains between the input and output lines.  
This plate can be removed if we maintain the same pitch for the signal lines as that for the ground plane \cite{Meander}.  
A QD is placed in the gap between the input and output lines.  
A similar design was previously tested using a CPS at a lower frequency \cite{Koppens06}.  
The simulations for the MW electric and magnetic fields are performed for our designs at 20 GHz using commercial software IE3D (Table \ref{table:simresults}).  
Figure \ref{fig:MWfield} shows the obtained in-plane electric field and perpendicular magnetic field profiles near the QD location.  
The input port is connected to the MW source and the output is grounded.  
The single line coil induces a stronger electrical field than expected, as discussed below.  
To concentrate the current density in the coil near the QD is a straightforward way to increase the magnetic field.  
This can be done by making the CPW line narrower toward the QD.  
However, in reality it also strongly increases the electric field in our simulation.  

Concerning the development of on-chip MW structures, there are already many studies performed on the on-chip spiral coil structure\cite{microcoil03}.  
We followed these studies and developed the on-chip spiral coil shown in Fig. \ref{fig:single}(b).  
We calculated the basic properties of our on-chip spiral coil and plotted them in Fig. \ref{fig:MWfield}.  
We adjusted the coil diameter to maximize the magnetic field.  
Compared with the single line structure, the electric field is greatly reduced but the magnetic field is small.  
Then we judge the spiral coil is not practical.  
When we calculate the ratio of the magnetic field to the electric field, we found that effectiveness is the same as that for the resonator type (see below).  
Note that regarding the field homogeneity, the on-chip spiral coil can produce the most homogeneous field.  

To minimize the electric field, we consider the on-chip resonator \cite{Blais04,Wallraff04} shown in Fig. \ref{fig:single}(c), which consists of a long CPW line of length $l$ with one port grounded.  
A standing wave with a voltage node at the ground port is forced along the line.  
The structure is similar to that of the single turn coil; however, the single turn coil has no resonance mode simply because the length is too short.  
If the coil has a resonance mode, it works more effectively at the resonant frequency.  
We calculate the effective wavelength $\lambda$ on the surface of GaAs to derive the resonance condition.  
Using the effective dielectric constant $\varepsilon' = (\varepsilon_r + 1)/2$, $\lambda$ is calculated as $\lambda_0 / \sqrt{\varepsilon '}$, where $\lambda_0$ is the wavelength in a vacuum.  

There are voltage and current nodes at different locations ($\pi/2$ phase shift).  
For $l = n \times \lambda / 2$ with an odd integer $n$, an antinode appears in the voltage at the QD location [at the center of the circle in Fig. \ref{fig:single}(c)].  
Under this condition, we expect a large electric field and a small magnetic field.  
Then $l$ is about 3 mm to satisfy the condition $l = \lambda / 2$ at 20 GHz, and therefore we insert a meander line\cite{Meander} to fit the line into a sample with a size of 2$\times$1.6 mm$^2$.  
When $n$ is an even integer, a node and an antinode appear in the voltage and the current, respectively.  
Therefore, the maximum magnetic field is expected at the QD location.  
The simulation results are plotted in Fig. \ref{fig:MWfield}.  
The data are obtained at 20 GHz both for comparison with the single turn coil and for the worst case calculation for ESR with a larger electric field.  
This resonator design clearly provides better characteristics for ESR than other designs and the strongest magnetic field.

\section{\label{sec:experiments}Experiments}

\subsection{\label{sec:cryostat}Experimental setup}
\begin{figure}
(a)\includegraphics[width=0.35\textwidth]{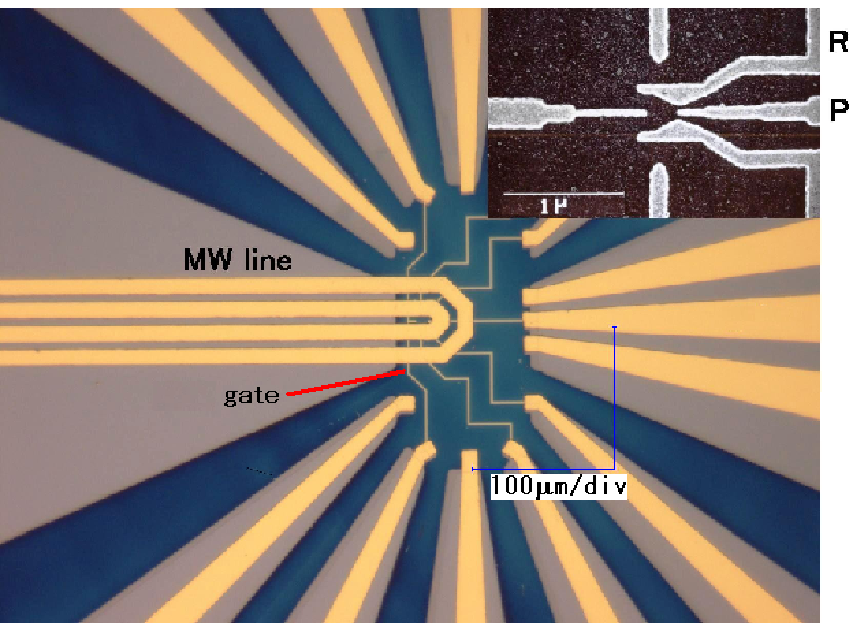}
(b)\includegraphics[width=0.35\textwidth]{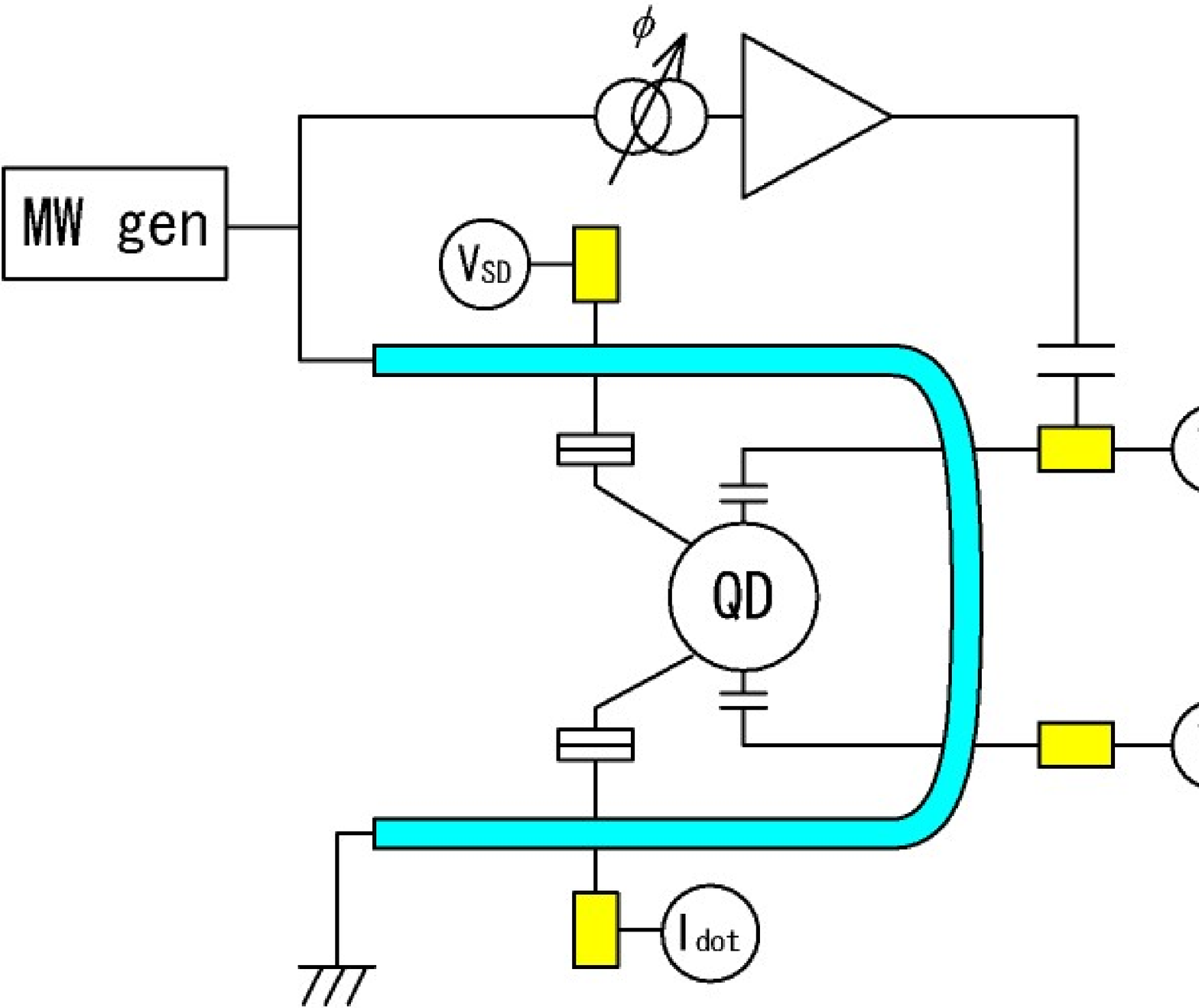}
\caption{
\label{fig:compcircuit}
(a) Photograph of the on-chip coil device including a QD structure in the center.  The inset gate metal pattern threading to the metal wires in the photograph is fabricated using electron beam lithography.  The other microwave metal line and metal plates are fabricated by optical lithography.  The microwave line is isolated from the gates by a 100 nm thick diluted photoresist layer.  We use a Canadian lateral dot design (Ref. \cite{Ciorga00}).  (b) Compensation circuit.  The triangle is a block attenuator.  
}
\end{figure}

We used our integrated on-chip coil and QD devices in a dilution refrigerator to characterize the MW performance.  
Two semirigid coaxial cables were fitted to supply the sample with MWs.  
These cables transmit signals of up to $\sim$46 GHz.  
The heat load was measured at $\sim$20 $\rm \mu$W per cable, which raised the base temperature to 40 mK.  
A gold CPW on an alumina substrate was connected to the semirigid coaxial cable via factory made glass beads.  
The high frequency port of the on-chip coil was connected to the alumina CPW with 1 mm long bonding wires.  

Images of the device and a circuit sketch are shown in Fig. \ref{fig:compcircuit}.  
We used a Canadian design \cite{Ciorga00} to make a lateral QD holding just a single electron and supplied a MW to the on-chip coil.  
We estimate that a 4 dBm input power produces 4 mA under impedance matching condition.  
At this power level the dilution refrigerator temperature increased to $\sim$ 300 mK.

\subsection{\label{sec:refandPAT}Reflection property and coulomb peaks}

\begin{figure}
\includegraphics[width=0.4\textwidth]{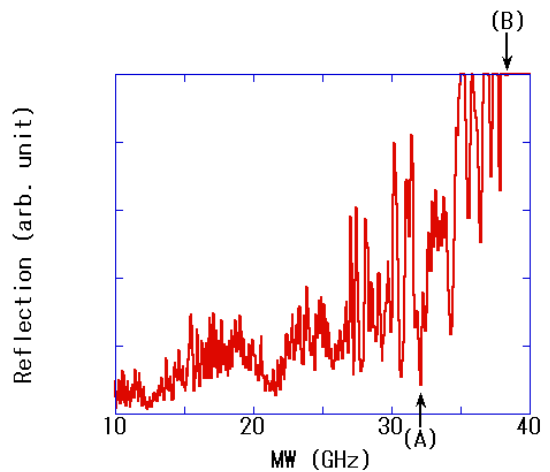}
\caption{\label{fig:reflection} 
Reflection coefficient as a function of microwave frequency.  The peaks and dips are anti-resonant and resonant points, respectively.  At the dip positions, high frequency signals can transmit more effectively.  Arrows (A) and (B) indicate the frequencies at which we measured the Coulomb peaks in Fig. \ref{fig:freqdepPAT}.  
}
\end{figure}

\begin{figure}
(a)\includegraphics[width=0.35\textwidth]{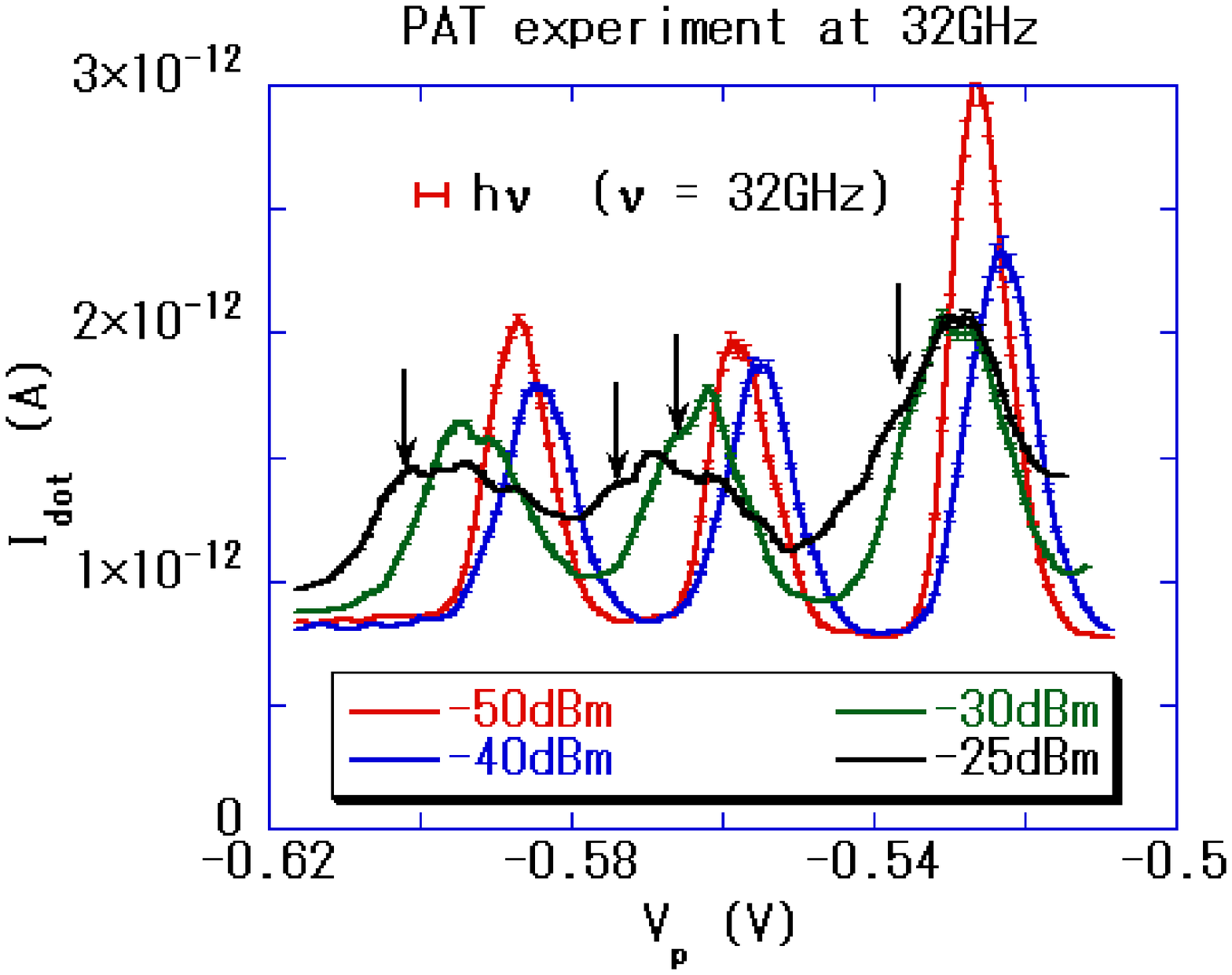}
(b)\includegraphics[width=0.35\textwidth]{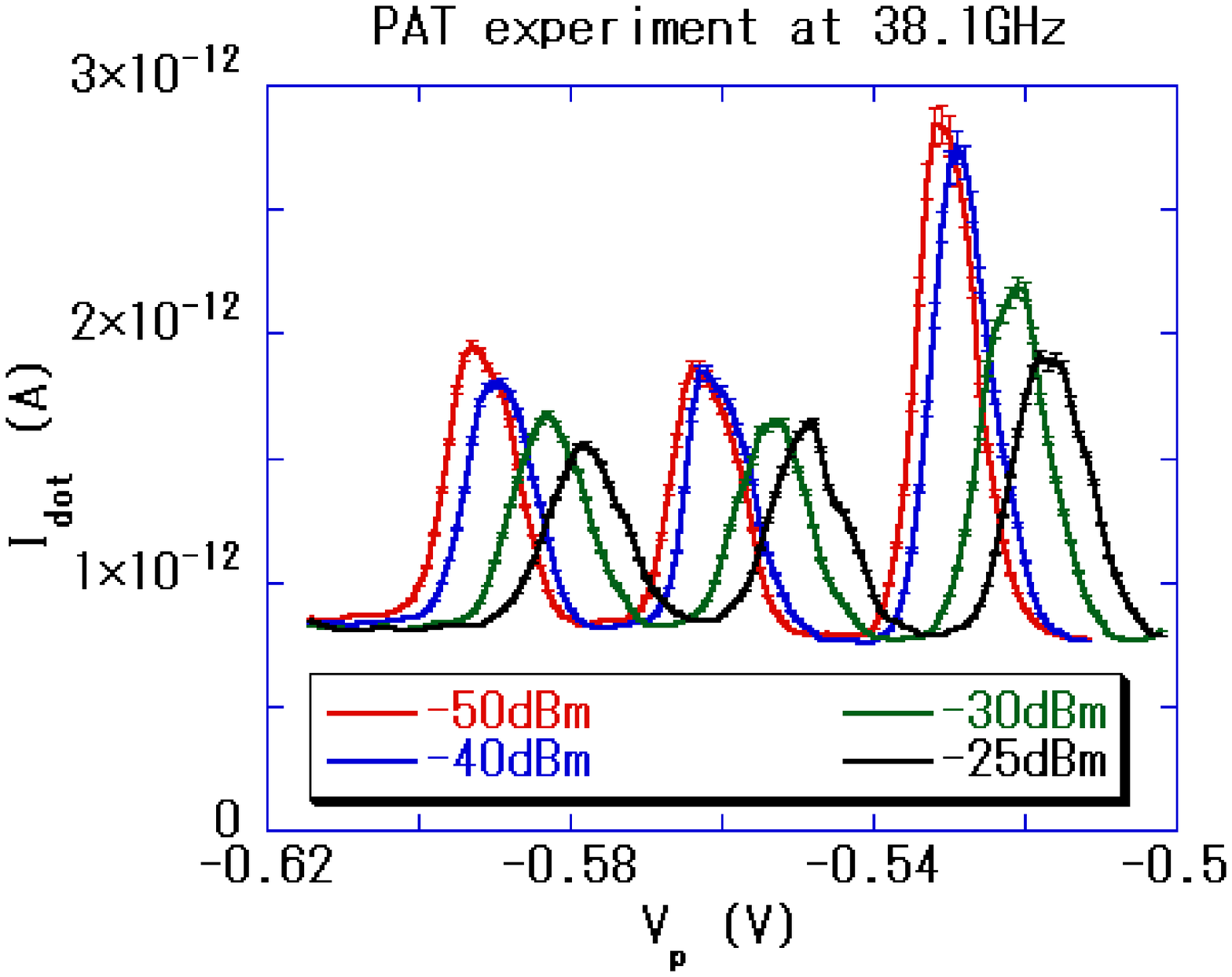}
\caption{\label{fig:freqdepPAT} 
Microwave modulation of Coulomb blockade spectrum measured at two different frequencies: MW reflection minimum (a) and maximum (b).  $I_{\rm dot}$ is the current flowing through the quantum dot.  Influence of the MW power on the Coulomb peaks are different between (a) and (b).  In (a) the peaks have side peaks (arrows) due to the photon assisted tunneling.  In (b) the main peaks shift and become broad without showing any clean PAT effect.  This behavior is due to the heating of the sample.  }
\end{figure}

In this section, we characterize the reflection coefficient of the coil circuit.  
For the first experiment we used a single turn coil similar to the design shown in Fig. \ref{fig:single}(a).  
With this coil, we expect a stronger PAT effect than for the resonator-type sample.  
When we applied a MW to the sample through the transmission line, part of the MW did not transmit.  
The reflection coefficient $\rho$ was measured using a network analyzer and a standing wave ratio bridge.  
The frequency dependence of the absolute value of $ \rho $ is shown in Fig. \ref{fig:reflection}, which we measured with an Anritsu autotester 560-98VF50A.  
There is a frequency region where the reflection is small.  
We can expect high efficiency for the MW performance at around this frequency region.  
Figure \ref{fig:freqdepPAT} shows the measured Coulomb peaks at two characteristic frequencies: (a) close to the local minimum (32 GHz) and (b) local maximum (38 GHz) of the reflection.  
The power dependence of the QD Coulomb blockade spectrum is shown in Fig. \ref{fig:freqdepPAT}.

Figure \ref{fig:freqdepPAT}(a) shows side peaks (arrows) caused by PAT and the peaks become broad with increasing MW power.  
On the other hand, Fig. \ref{fig:freqdepPAT}(b) shows the principal peaks just shifting with increasing MW power for the frequency at the antiresonance value of 38 GHz.  
The MW is not transmitted into the sample.  
Then Joule heating along the coaxial line causes the peak to drift.  
The MW at the resonance frequency propagates to the on-chip coil and generates the PAT signal.  
These results provide a guideline about the MW frequency where we should work for the ESR measurement.  
Nevertheless, the PAT process still presents a problem.  
The next step is to focus on applying a magnetic field and reducing the electric field in order to remove the PAT signal.

\subsection{\label{sec:compensation}PAT compensation}

\begin{figure}
\includegraphics[width=0.40\textwidth]{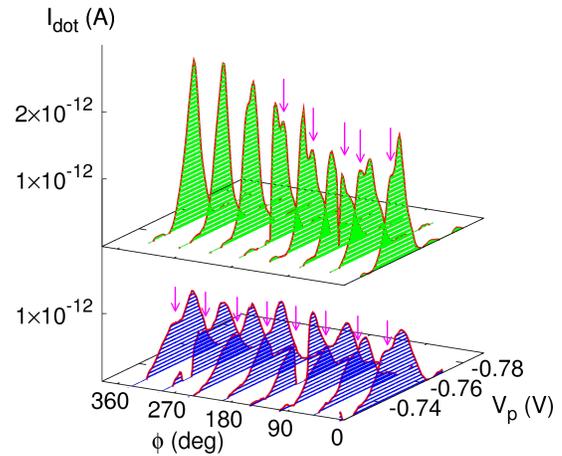}
\caption{\label{fig:compensation} 
Dependence of Coulomb peak on compensation $\phi$.  The MW frequency is 40 GHz and the MW power is -23 (-20) dBm for the upper (lower) plot to the MW line.  The upper (lower) trace is for -10 (-23) dBm compensation at gate R.  The horizontal plane is the $V_p - \phi$ plane.  In the upper trace, the compensation works well and the peak reaches its maximum at $\phi$ = 288$^\circ$.  
}
\end{figure}

To detect ESR, it is desirable to attenuate the PAT process.  
We developed the compensation circuit \cite{Koppens06} to operate in the more active way and at the much higher frequency.  
We used the resonator-type on-chip coil, as shown in Fig. \ref{fig:compcircuit}.  
We split the MW signal and used a block attenuator to adjust the compensation level.  
After adjusting the phase $\phi$ with a mechanical phase shifter (Waka Manufacturing Co., Ltd.) to maximize the compensation level, we launched the MW into one of the side gates through a bias tee A3N1025 (Anritsu Corp).  

The compensation results are shown in Fig. \ref{fig:compensation}.  
The MW frequency is 40 GHz.  
If the compensation level matches the electric field provided by the on-chip coil, the side peak (arrow) is suppressed and the main peak is increased.  
This is the case shown in the upper panel of Fig. \ref{fig:compensation}.  
The main peak becomes sharpest when the phase is best adjusted.  
When the compensation level does not match, the compensation works poorly as seen in the lower panel of Fig. \ref{fig:compensation}.  
Under the best compensation condition, there is only a standing magnetic field, which is a good regime for detecting ESR.   
The active compensation technique holds for any frequencies, using a continuous phase shifter.  
Actually we obtained similar results at other frequencies.  
When the phase matches and compensation works well, we can apply almost $\sim$ 0 dBm at the sample edge.

\section{\label{sec:discussion}DISCUSSION}

\begin{figure}
(a)\includegraphics[width=0.35\textwidth]{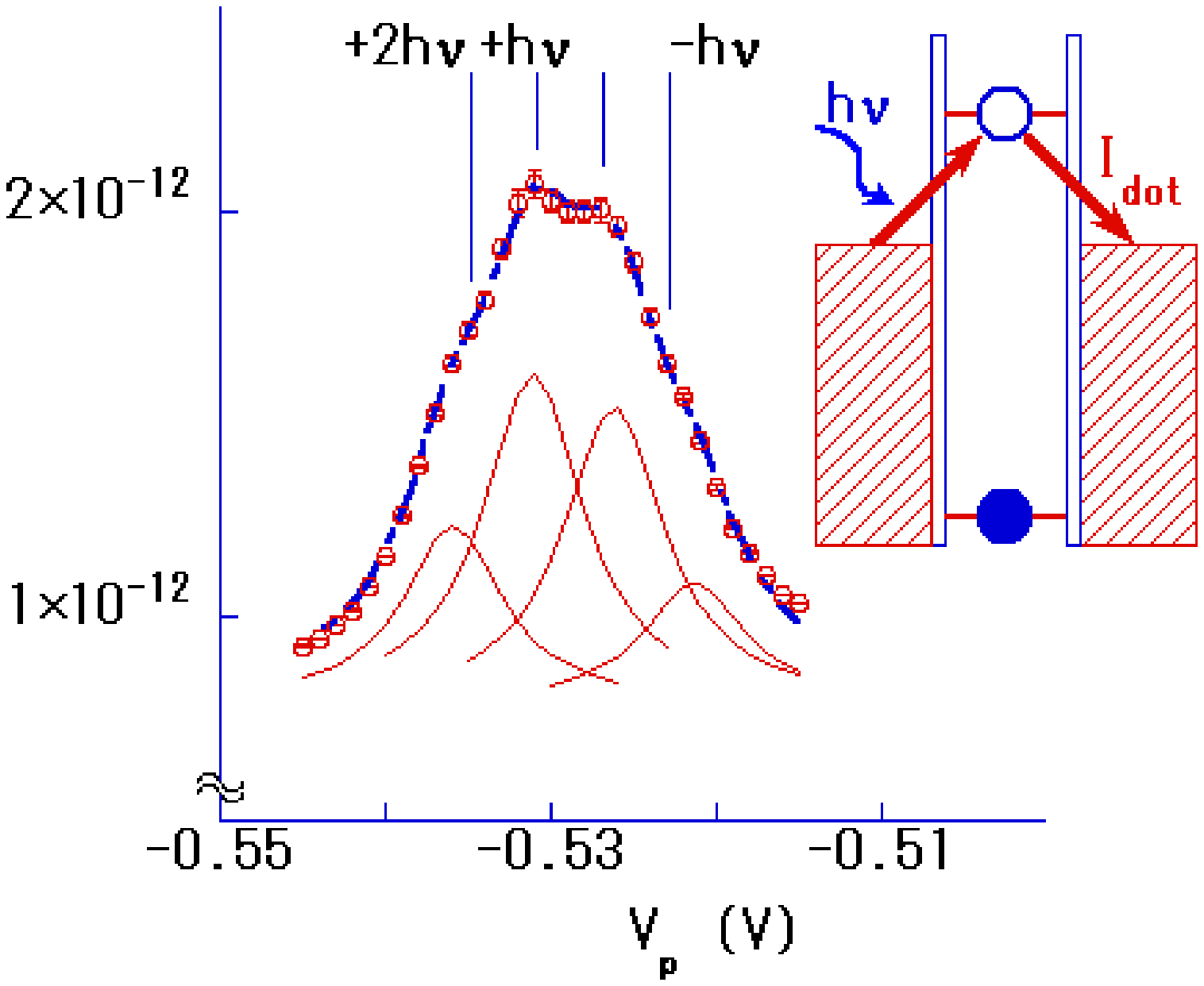}
(b)\includegraphics[width=0.35\textwidth]{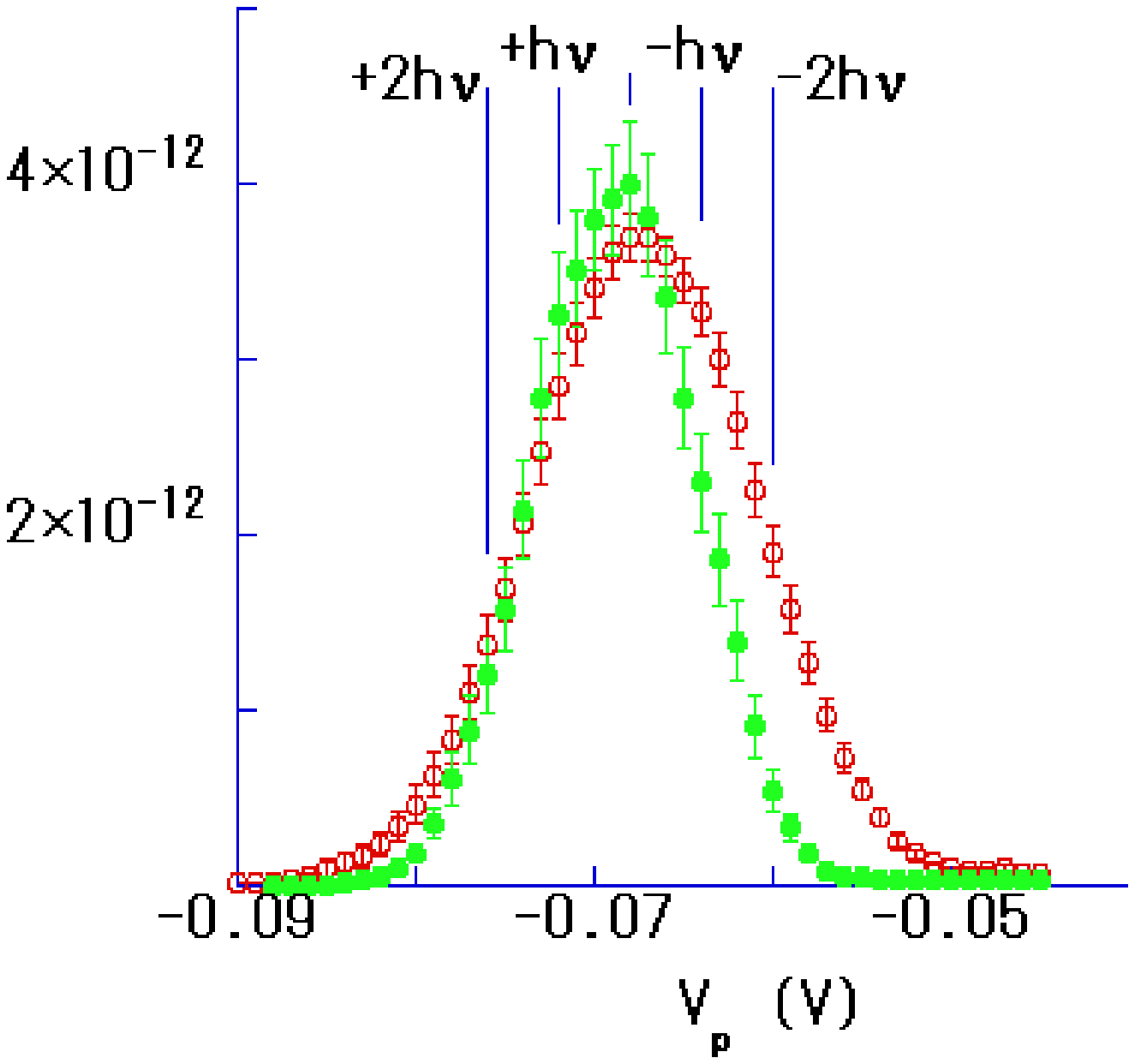}
\caption{\label{fig:fitting}  Lorentzian fit to the PAT signal.  (a) Data with the single turn coil MW line in Fig. \ref{fig:freqdepPAT}(a).  MW frequency, and power is 32 GHz, and -30 dBm, respectively.  (b) Data for comparison with the on-chip resonator type MW line.  MW frequency is 28 GHz.  Open and solid circles are for the MW power of -30 and -50 dBm, respectively.  The experiment is performed for the lowest reflection.  Therefore, the MW frequencies are slightly different between (a) and (b).  
}
\end{figure}

From both the simulation and experiment, we confirmed that the resonator-type coil is a better choice.  
To characterize the coil performance on the QD more quantitatively, we now analyze the effect of MW modulation on the Coulomb blockade peaks in Figs. \ref{fig:freqdepPAT} and \ref{fig:compensation}.  
The change in the peak arises from the PAT process such that electrons in the QD are injected (ejected) from (to) the reservoirs absorbing (emitting) photons \cite{Kouwenhoven94PAT,WilfredPATrev,Wilfred03}.  
Due to the PAT process there appear several small peaks/shoulders (arrows) with the spacing given by the photon energy $h\nu$.  
Following the standard PAT model, we define an oscillating potential $\tilde V \cos(2 \pi \nu t)$ at the QD.  
A voltage drop across a tunnel barrier modifies the tunnel rate through the barrier as \cite{PATtunneling}
\[
\tilde \Gamma (E) = \sum\limits_{n =  - \infty }^\infty  {J_n ^2 (\alpha ) \cdot \Gamma (E + n h\nu )}.  
\]
Here $n = 0, \pm 1, \pm 2, ...$.  
$\tilde \Gamma (E)$ and $\Gamma (E)$ are the tunnel rates at energy $E$ with and without MW irradiation, respectively.  
$J_n$'s are Bessel functions of the first kind.  
We define $\alpha = e \tilde V / h \nu$.  
We apply this equation to the simplest case of a nondegenerate single QD level.  
We focus on the situation that the electron number changes from $N-1$ to $N$ (see the inset in Fig. \ref{fig:fitting}).
We take into account the charge balance between the QD and source-drain and the electron number conservation law (the so-called master equations) and calculate the net current as  
\begin{equation}
\label{eq:PAT}
 I_{dot} = \frac{{e\gamma _l \gamma _r }}{{\gamma _l  + \gamma _r }}\frac{b V_{SD}}{k_B T} \sum\limits_{n =  - \infty }^\infty  {J_n ^2 (\alpha )\left. {\frac{{\partial f(x)}}{{\partial x}}} \right|_{x = \frac{E_N - aV_p  - nh\nu}{k_B T}} }, 
\end{equation}
$E_N$, $\gamma_{l(r)}$, $V_p$, and $V_{SD}$ are the energy level of dot, bare tunnel rate to the left (right) contact, gate voltage, and source-drain bias voltage, respectively.  
$f(x)$ is the Fermi-Dirac distribution function and coefficients $a$ and $b$ are the lever arms measured by nonlinear Coulomb blockade spectroscopy \cite{Kouwenhoven_CImodel}.  
$ {\partial _x f(x)} _{x = ( E_N - aV_p  - nh\nu ) /k_B T} $ gives a peak at $a V_p = E_N - n h \nu$ with amplitude proportional to $J_n^2(\alpha)$.  

The electric field produced by the coils can be estimated by fitting the line shape of the Coulomb blockade peak to Eq. \ref{eq:PAT}.  
We approximate $\partial f(x)/\partial x$ by a Lorentzian function because the data contain many kinds of experimental errors and take into account up to the two-photon process.  
For the single turn coil, we analyzed the data at 32 GHz with -30 dBm input (4.5 mV at the sample edge) in Fig. \ref{fig:freqdepPAT} because the data show clear PAT structures.  
One of the fitting results is shown in Fig. \ref{fig:fitting}(a).  
The fitting parameters have some ambiguity, but we finally estimated the averaged voltage drop $\tilde V$ of 150 $\rm\mu$V using the value of $\alpha = 1.2$.  
The distance $d$ crossing the tunnel barrier was estimated to be 300 nm yielding an electric field of $E = \tilde V / d \sim$  0.5 mV/$\rm\mu m$.  

For the resonator type, we measured the MW modulation under the similar condition (at 28 GHz with -30 dBm input) for comparison in Fig. \ref{fig:fitting}(b).  
We tried but the PAT structure was not well resolved.  
We could only estimate $\alpha = 0.45$ from the change of the main peak to be proportional to $J_0^2(\alpha)$ and the electric field of 0.17 mV/$\rm\mu m$ at 28 GHz with -30 dBm input.  
The MW modulation at 40GHz shown in Fig. \ref{fig:compensation} was more easily analyzed.  
In the lower panel, we estimated 0.6 mV/$\rm\mu m$ with -20 dBm input by fitting.  
Then we estimated that the electric field at 40 GHz would be 0.2 mV/$\rm\mu m$ with -30 dBm input.  

The single turn coil data show larger MW modulation than that for the resonator coil in both simulation and experiment.  
The simulation generally gives us a good guideline for designing the on-chip coil.  
The electric field experimentally evaluated is, however, several times larger than the simulation.  
Another problem is that the simulation for the resonator type predicts magnetic resonance at 40 GHz with zero electric field, whereas the experiment shows a finite electric field under the resonance condition.  
We consider that these problems come from the bonding wires for connecting the input and ground because they can produce an extra impedance and modulate the resonance mode.  
The misalignment between the MW lines and the QD also gives a larger electric field in the experiment than that in the simulation because the MW lines are symmetric around the gap where the QD is placed.  
The alignment precision is restricted by an optical mask aligner precision, which is a few micrometers for the high frequency lines.  

To estimate the MW magnetic field, we compared the electric field for PAT data to the simulation and assumed that the ratio between the MW electric and magnetic fields is the same for the simulation.  
The estimated magnetic field is $\sim$ 0.4 g at -30 dBm.  
We used the ratio in the case that electric field is larger (see Sec. \ref{sec:simulation}) and thus this estimation is somewhat underestimation of MW magnetic field.  
Under the compensation condition described above, we can apply a 0 dBm input, and then the magnetic field will be 30 times larger or $\sim$ 1 mT.  
Note with standard cavity resonators, it is very difficult to produce such a high magnetic field.  
If electron $g$ factor is 0.35 \cite{Koppens06}, the spin flip time is 2.5 MHz, which is detectable level when the tunneling rate of Coulomb blockade peak is about 10 MHz (1pA current).  
We can expect to detect an ESR signal under these conditions.

\section{\label{sec:conclusion}Conclusion}
We reported designs for a MW band on-chip coil and experimental techniques for actual devices with QDs.  
We simulated three types of on-chip coil structure.  
The single turn coil pattern is the simplest but it produces the strongest MW electric field in our models.  
In accordance with conventional NMR and ESR studies\cite{microcoil03}, we designed an on-chip spiral coil.  
This produces a homogeneous magnetic field but a smaller amplitude.  
Another disadvantage for the spiral coil is that more microfabrication steps are necessary, and therefore the production yield will be low.  
Finally, the resonator-type pattern appeared more relevant than the others.  
We then fabricated samples with the resonator-type on-chip structure.  
We examined experimentally how strongly impedance matching affects the Coulomb blockade transport.  
With a low reflection coefficient, the current was modulated by a high frequency signal and there significantly appeared the PAT effect.  
It is clear that this PAT process overrides the electron spin resonance signal.  
To suppress the PAT effect and maximize the MW magnetic field, we developed a compensation technique available for the MW band.  
We demonstrated that the PAT side peak disappeared when the phase of the compensation signal was best adjusted.  
According to our calculation, we could predict that a sufficiently strong MW magnetic field is generated for the single ESR measurement with QD.

\section{Acknowledgments}

The authors thank W.G. van der Wiel for fruitful discussions.  
S.T. acknowledges financial supports from the Grant-in-Aid for Scientific Research A (No. 40302799), and B (No. 18340081), SORST Interacting Carrier Electronics, JST, and Special Coordination Funds for Promoting Science and Technology, MEXT.  

\newpage 

\end{document}